\documentclass[apj]{emulateapj}

\newcommand {\apgt} {\ {\raise-.5ex\hbox{$\buildrel>\over\sim$}}\ }
\newcommand {\aplt} {\ {\raise-.5ex\hbox{$\buildrel<\over\sim$}}\ }

\shorttitle{The formation of star clusters through mergers}
\shortauthors{Fujii et al.}

\begin{document}


\title{The formation of young dense star clusters through mergers}


\author{M. S. Fujii\altaffilmark{1}, T. R. Saitoh\altaffilmark{2}, 
and S. F. Portegies Zwart\altaffilmark{1}}
\altaffiltext{1}{Leiden Observatory, Leiden University, NL-2300RA Leiden, The Netherlands}
\email{fujii@strw.leidenuniv.nl}
\altaffiltext{2}{Interactive Research Center of Science, Tokyo Institute of Technology, 2-12-1 Ookayama, Meguro, Tokyo 152-8551, Japan}



\begin{abstract}

Young star clusters like R136 in the Large Magellanic Cloud and NGC
3603, Westerlund 1, and 2 in the Milky Way are dynamically more
evolved than expected based on their current relaxation times. In
particular, the combination of a high degree of mass segregation, a
relatively low central density, and the large number of massive
runaway stars in their vicinity are hard to explain with the
monolithic formation of these clusters. Young star clusters can
achieve such a mature dynamical state if they formed through the
mergers of a number of less massive clusters. The shorter relaxation
times of less massive clusters cause them to dynamically evolve
further by the time they merge, and the merger product preserves the
memory of the dynamical evolution of its constituent clusters. With a
series of $N$-body simulations, we study the dynamical evolution of
single massive clusters and those that are assembled through merging
smaller clusters together.  We find that the formation of massive star
clusters through the mergers of smaller clusters can reproduce the
currently observed spatial distribution of massive stars, the density,
and the characteristics (number and mass distribution) of the stars
ejected as runaways from young dense clusters. We therefore conclude
that these clusters and possibly other young massive star clusters
formed through the mergers of smaller clusters.

\end{abstract}



\section{Introduction}
Young ($\aplt 4$ Myr) and dense ($\rho \apgt 10^4\,$$M_{\odot}$/pc$^3$) star
clusters, NGC 3603, Westerlund 1 and 2 in the Milky Way Galaxy and R136
in the Large Magellanic Cloud (LMC), pose a number of interesting
evolutionary problems on the formation and early evolution of star
clusters.  Their relatively small number of stars and their young ages
make them ideal modeling targets \citep[see][]{2010ARA&A..48..431P},
because they can be simulated on a star-by-star basis for their entire
lifetime. In addition, the observational data on these clusters have
excellent quality, which makes them ideally suited for comparison
with the results of the numerical simulations.

From an observational point of view, each of these clusters shows a
remarkably mature dynamical state compared to their ages
\citep{2008ApJ...675.1319H,2009ApJ...707.1347A,2007A&A...466..137A,
2011MNRAS.412.2469G}; the degree of mass segregation suggests that
either these clusters are born with a certain degree of mass
segregation or that it develops more quickly than expected by 
standard relaxation theories.

In particular, the LMC cluster, R136, shows characteristics of a
post-core-collapse star cluster \citep[][hereafter
  FPZ2011]{2003MNRAS.338...85M,2011Sci...334.1380F}.  The cuspy
density profile of R136 \citep{2003MNRAS.338...85M} and the large
number of massive runaway stars which seem to have escape from R136,
such as VFTS 682 \citep{2011A&A...530L..14B}, 30 Dor 016
\citep{2010ApJ...715L..74E} and several OB stars
\citep{2007ASPC..367..629B,2010A&A...519A..33G}, are effective
signatures of core collapse. The mass and number distribution of the
runaway stars ejected from R136 can be explained by three-body
interactions between a hard binary and a single star
\citep[FPZ2011;][]{2011MNRAS.410..304G,2012ApJ...746...15B}.  Such a
``bully'' binary (BB) that formed in the gravothermal collapse of the
cluster kicks out its surrounding stars and produces runaway stars.
In FPZ2011, we demonstrated that the number of runaway stars that one
BB can produce does not depend on the global parameters of the cluster, 
and accounts for $\sim 21$ in
total. The fraction of runaway stars to the whole cluster, $f_{\rm
  run}$, is therefore larger in clusters with fewer stars.  The
observed fraction of runaway stars is satisfactorily explained by the
cluster having experienced core collapse within 1 Myr since its birth
(see FPZ2011).

Such an early core collapse can be initiated sufficiently quickly if
the initial density of the cluster core $\rho_{\rm c} \apgt 10^6
M_{\odot} \mathrm{pc}^{-3}$, which is considerably higher than the
currently density observed in R136: $\rho_{\rm c} \sim 5\times 10^4
M_{\odot} \mathrm{pc}^{-3}$ \citep{2003MNRAS.338...85M}.  With a
current half-mass relaxation time of $t_{\rm rh}\sim 100$ Myr
\citep{2003MNRAS.338...85M}, it is unlikely that R136 evolved to a
state of core collapse within 1\,Myr and subsequently evolved towards
the less dense state it is in today.

One way to speed up the initial dynamical evolution towards a state of
core collapse is by forming the cluster from the hierarchical merging
of several smaller clusters \citep{2007ApJ...655L..45M}. The shorter
relaxation timescales of these smaller clusters cause them to evolve
dynamically more quickly than one single large cluster
\citep{1972A&A....21..255A,2007ApJ...655L..45M,2009MNRAS.400..657M,
  2011ApJ...732...16Y}. A merger between these dynamically evolved
clusters preserves the memory of their past dynamical evolution and
imprints this onto the more massive post-merger clusters
\citep{2007ApJ...655L..45M}.

Sub-clustered star formation seems to be common for young stars and
protostars embedded in molecular clouds
\citep{2009ApJS..184...18G,2003ARA&A..41...57L,2010ARA&A..48..431P}. Numerical
simulations of star formation in turbulent molecular clouds also
support star cluster formation via hierarchical mergers
\citep{2003MNRAS.343..413B,2004MNRAS.349..735B,2011MNRAS.410.2339B,
2011MNRAS.413.2741G,2012MNRAS.419..841K}. Not only for young
clusters in the Galactic disk, the merger scenario is suggested for
extended old globular clusters in the Milky Way-like NGC2419
\citep{2011ApJ...729...69B,2011A&A...529A.138B} and massive star
clusters which are born during galaxy mergers
\citep{2011IAUS..270..483S}.  Thus, the formation of star clusters via
hierarchical merging of sub-clusters might be very common for any
kinds of star clusters.

The hierarchical merging of small sub-star clusters solves two
problems in our understanding of R136 and other clusters with a
similar age, density, mass, and mass function: (1) It allows for the
cluster with a long relaxation time to experience core collapse well
within the time frame for these sub-clusters, and (2) because each of
the sub-clusters experiences an individual core collapse before they
merge, the number of runaway stars produced by dynamical sling shots is
considerably higher.  Since the number of runaways is independent of
cluster mass (FPZ2011), the expected number of O-type runaway stars is
only one or two for R136 if it formed as a single cluster, but at
least three were observed
\citep{2010A&A...519A..33G,2010ApJ...715L..74E}.  It is therefore
surprising that R136 has produced more runaways than anticipated by
this theory if it was born as a single cluster.  The hierarchical
merging formation of R136 solves this conundrum.

The hierarchical merging of star clusters is also preferable for the
formation of very massive ($>150 M_{\odot}$) stars found in R136 and
NGC 3603 \citep{2010MNRAS.408..731C}.  Runaway collisions of stars
during core collapse easily form such very massive stars in the
cluster core \citep{1999A&A...348..117P,2009ApJ...695.1421F}.  These
massive stars stay in the cluster center due to mass segregation and
form BBs. In this way, each sub-cluster that experiences
core collapse will lead to the formation of one BB. After the 
sub-clusters merged, the massive BBs interact with each other in the
core of the merged cluster. The binary-binary interactions can form a
massive escaping binary like R145
\citep{2009MNRAS.395..823S,2011MNRAS.410..304G}.
 
We also argue that Westerlund 1 and 2 are still in the
hierarchical merging process, which is demonstrated by their clumpy
appearance. These clusters have not completed their merging processes, 
which has profound consequences for their current degree of
mass segregation and the distribution of their stars
\citep{2009Ap&SS.324..321M,2011MNRAS.412.2469G,2011MNRAS.416..501R}.

In this paper, we demonstrate that merging star clusters can explain
the dynamically evolved characteristics of young dense clusters in the
Milky Way and the LMC.  The observed distributions of massive stars
including stars with an initial mass of $\apgt 150 M_{\odot}$, such as
those found in R136 and NGC 3603, which are consistent with our
results.  Furthermore, the runaway stars around R136 and
Westerlund 2 are consistent with our merger models. The core collapse
in each small sub-cluster before merger forms a massive hard binary in
its core. These massive binaries in the sub-clusters produce a
sufficient amount of runaway stars, and then they are ionized or
ejected from the cluster after their hosts merged. We demonstrate that
such massive stars exhibit a spatial distribution similar to those
observed in the region of R136 and NGC 3603.  These dynamical
processes result in a relatively low density of the cluster compared
to their initial density or the density during the core collapse; the
core density of these merger remnants in our simulations is consistent
with those of observed young clusters.

\section{Methods}
We performed a series of $N$-body simulations of single and multiple
star clusters that merge to form a large single cluster.  For merger
cases, we adopted a mass of $6.3\times 10^3M_{\odot}$ and a King model
\citep{1966AJ.....71...64K} with the dimensionless central potential
$W_0=2$ for each sub-cluster. The initial mass function (IMF) of stars 
were
drawn from the Salpeter mass function \citep{1955ApJ...121..161S}
between 1 and 100 $M_{\odot}$. We imposed that each cluster contains
at least one star with $m> 80 M_{\odot}$. The initial half-mass radius
and core density of our simulated clusters are $r_{\rm h} \simeq 0.1$
pc and $\rho_{\rm c} \simeq 2\times 10^6$ $M_{\odot}$ pc$^{-3}$,
respectively.  This core density is similar to the central density of 
sub-clusters formed in star formation simulations in a turbulent 
molecular cloud \citep{2004MNRAS.349..735B}.  With these
initial conditions our model clusters have $N=2048$ (2k) stars and a
half-mass relaxation time of $t_{\rm rh} \simeq 0.37$\,Myr.  We call
this model single-2k. The parameters are summarized in table
\ref{tb:model}.

We initially distribute 4 or 8 of these sub-clusters (single-2k) 
randomly in a volume with a radius of $r_{\rm max}$ and with zero
velocity.  We varied $r_{\rm max}$ from 1\,pc to 6\,pc.  The
clusters merge within a few Myr to a single cluster. The total masses
of the merger remnant correspond to the mass of NGC 3603 for 4-cluster 
mergers and R136 for the 8-cluster merger cases. We summarize the 
initial conditions in table \ref{tb:init}.

For comparison, we also performed simulations of single clusters,
whose masses are identical to R136.  For this single cluster model, we
adopted a King model for the initial density profile with $W_0=6$; we
call this model single-w6. The core density of model single-w6 is
comparable to those of the individual single-2k models, but with a
total mass $M \simeq 5\times 10^4 M_{\odot}$ and a half-mass radius
$r_{\rm h} \simeq 0.3$ pc.  The resulting half-mass relaxation time,
$t_{\rm rh}$, is $\sim 4.4$\,Myr, which enables these clusters to
experience collapse within $\sim 1$ Myr
\citep{2002ApJ...576..899P,2004ApJ...604..632G}.

The models, single-2k and single-w6, have a higher initial central
density than those of observed star clusters. This choice of initial
density is motivated by our requirement that the cluster should
experience core collapse within $\sim 1$ Myr, which is required for
the clusters to effectively to produce runaway stars. However, the
core density drops and becomes comparable to the observed ones after
the core collapse due to the scattering of stars in the cluster
center. We discuss more details in section \ref{sec:mass_dist}.

We additionally performed simulations of a single model with the same 
parameters as single-w6 but with a smaller lower-mass limit of the IMF. 
For our standard models, we adopted 1 $M_{\odot}$ as a lower-mass limit, 
which is more massive than observed ones. We adopted 0.47$M_{\odot}$ as a 
lower-mass limit for this model (single-lm), and as a consequence the mean 
mass of single-lm is half of single-w6. 

Furthermore,
we performed two additional series of calculations of single clusters
with the same mass and IMF as for the single-w6
model.  One of these models has $W_0=8$ and half-mass radius 
$r_{\rm h} \simeq 1.4$ pc; we call this model single-w8.  This model 
initially has a density profile similar to R136.  The core density 
for this model is $1.6\times 10^5$ $M_{\odot}$pc$^{-3}$ and its 
relaxation time is $\sim 43$ Myr; this cluster is therefore not 
expected to experience core collapse within $\sim 3$\,Myr
\citep{2002ApJ...576..899P,2004ApJ...604..632G}.  The other model is
with primordial binaries; which we name single-pb.  We assigned the
most massive 3\% of stars ($\apgt 8 M_{\odot}$) as equal-mass
primordial binaries with the binding energy of 3 kT, which corresponds
to an orbital period of $\sim 3\times10^3$--$10^6$~days.  Here kT is
the fundamental unit of kinetic energy and 1.5 kT corresponds to the
average kinetic energy of stars in the cluster before we add
binaries. Binaries with this binding energy have intermediate hardness
and therefore interact efficiently with single stars and with each
other \citep{2009PASJ...61..721T} and therefore they are expected to produce
runaway stars, while binaries with shorter orbital periods are 
dynamically inactive due to their small cross sections (FPZ2011).  
We choose equal-mass binaries because
massive stars observationally prefer massive companions
\citep{2011IAUS..272..474S,2007ARA&A..45..481Z}.  All binaries are
initialized with circular orbits and random orbital phase and
inclination.  In table \ref{tb:init} we summarize the initial
conditions for these models.

During our simulations, we applied mass loss due to stellar winds only
for stars more than $100M_{\odot}$.
Initially our simulations do not contain such massive stars, but they
form occasionally due to collisions, for which we adopted the
sticky-sphere approach. For the rate of mass loss by the stellar wind,
for stars with $m>100 M_\odot$, we adopted $\dot{m}=5.7\times
10^{-8}(m/M_{\odot})$ $(M_{\odot}\, \rm{yr}^{-1})$
\citep{2009ApJ...695.1421F}.  The stellar radii were adopted to have
zero-age main-sequence radii for solar metalicity
\citep{2000MNRAS.315..543H}.

\begin{figure*}[htbp]
\begin{center}
\epsscale{1.00}
\plottwo{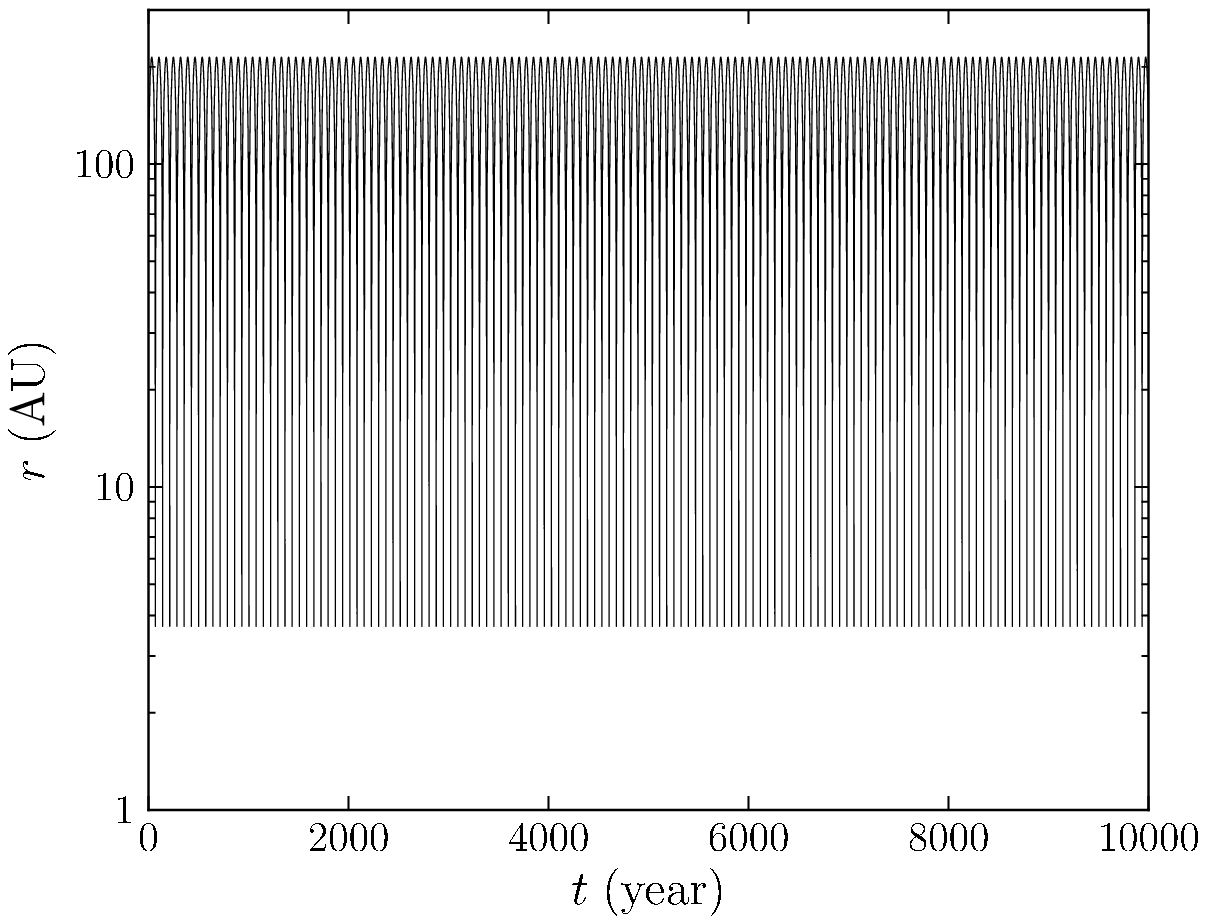}{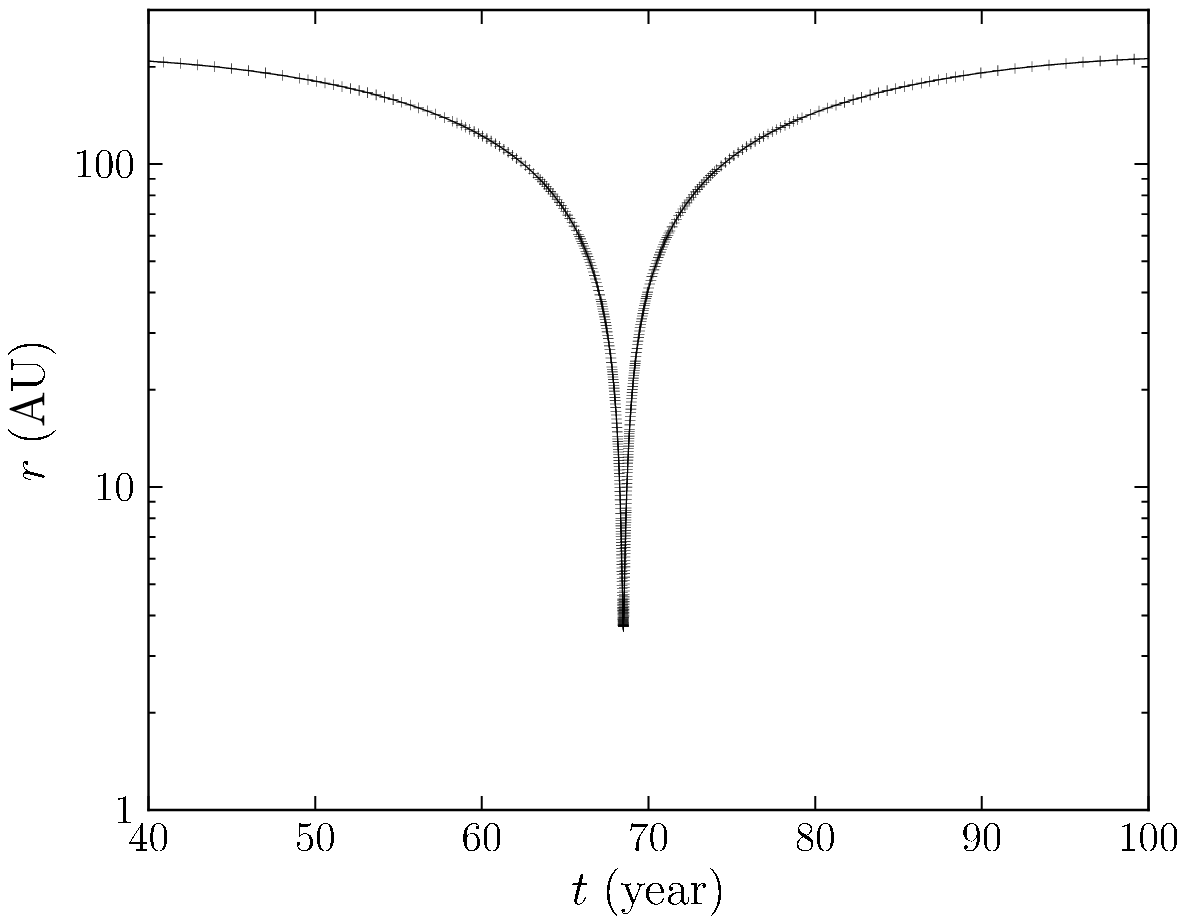}
\caption{Orbital separation of a hard binary as a function of time (right for 0.1 Myr and left: for one orbital period). The masses of the binary components are 100 and 150 $M_{\odot}$, respectively. The orbital period, semi-major axis, and eccentricity are 72 year, 109 AU, and 0.966, respectively. In right panel, crosses indicate the time steps adopted by our timestep criterion \citep{2008NewA...13..498N} with an accuracy parameter of 0.15. The minimum and maximum stepsizes in this period are 0.73 day and 1.0 year, respectively.
\label{fig:two_body}}
\end{center}
\end{figure*}

Each simulation was performed up to an age of 3\,Myr, which is the
time for the first supernova explosion.  The equations of motion were
integrated using a sixth-order Hermite scheme with individual
timesteps with an accuracy parameter $\eta =$ 0.15 -- 0.3
\citep{2008NewA...13..498N}.  We adopted the accuracy parameter to
balance speed and accuracy.  Our code does not include any special
treatment for hard binaries.  The sixth-order Hermite scheme, however,
can deal with hard binaries which form in our simulations. We tested
our code integrating hard binaries with an eccentricity of
$>0.9$. Figure \ref{fig:two_body} shows an orbital separation of a
binary. The masses of the binary components are 100 and 150
$M_{\odot}$, respectively, and their orbital period, semi-major axis,
and eccentricity are 72 year, 109 AU, and 0.96. This binary has the
same parameters as one of hard binaries which actually formed in our
simulations. The energy and angular momentum errors maintain
$<10^{-6}$ and $<2\times 10^{-8}$ in 0.1 Myr. Even if the energy error
increases linearly, the relative error in the energy is conserved to
less than 0.1\% over the 3 Myr of the simulation: our code treats
close stellar encounters down to $\sim 1$ AU accurately.  Stars that
approach each other within 1 AU generally participate in a collision
and therefore such close encounters do not need to be resolved.  For
each simulation we confirmed that the energy error of clusters is
below 0.1\% of the initial total energy of the clusters.

\begin{table*}[htbp]
\begin{center}
\caption{Single cluster models.\label{tb:model}}
\begin{tabular}{cccccccccc}\hline
Model & $N$ & $M$ ($M_{\odot}$) & $W_0$ & $r_{\rm h}$ (pc) & $\rho_{\rm c}$ ($M_{\odot}$pc$^{-3}$) & $\sigma$ (km/s) & $t_{\rm rh}$ (Myr) & $m_{\rm min}$ ($M_{\odot}$) & $N_{\rm run}$\\ \hline\hline
single-2k & 2048 & $6.3\times 10^3$ & 2 & 0.097  & $2.2\times 10^6$ & 11 & 0.37 & 1.0 & 7\\ 
single-w6 & 16384 & $5.1\times 10^4$ & 6 & 0.32 & $1.7\times 10^6$ & 17 & 4.4 & 1.0 & 6\\
single-lm & 32768 & $5.1\times 10^4$ & 6 & 0.32 & $1.7\times 10^6$ & 17 & 8.3 & 0.47 & 3\\
single-w8 & 16384 & $5.1\times 10^4$ & 8 & 1.4  & $1.6\times 10^5$ & 6 & 43 & 1.0 & 1\\ 
single-pb & 16875 & $5.1\times 10^4$ & 6 & 0.32 & $1.7\times 10^6$ & 17 & 4.4 & 1.0 & 1\\ \hline
\end{tabular}
\end{center}
\end{table*}

\begin{table*}[htbp]
\begin{center}
\caption{Merger models.\label{tb:init}}
\begin{tabular}{cccc}\hline
Model & $N_{\rm cl}$\tablenotemark{a} & $r_{\rm max}$ (pc)\tablenotemark{b} & $N_{\rm run}$\tablenotemark{c} \\ \hline\hline
m2k8r1 & 8 & 1  & 1 \\ 
m2k8r3 & 8 & 3  & 1 \\ 
m2k8r5 & 8 & 5  & 2 \\ 
m2k8r6 & 8 & 6  & 2 \\ 
m2k4r3  & 4 & 3 & 3  \\ \hline
\end{tabular}

\tablenotetext{1}{The number of sub-clusters.}
\tablenotetext{2}{The radius in which sub-clusters initially distribute.}
\tablenotetext{3}{The number of runs which we performed.}
\tablecomments{The models are named according to the following rules; ``m'' indicates merger models, the name of models for the sub-clusters, the number of sub-clusters, and the value of $r_{\rm max}$. For example, m2k8r1 stands for a merger model of eight single-2k clusters which are located within $r_{\rm max}=1$ pc.}
\end{center}
\end{table*}

\section{Comparison between simulations and observations}

\begin{figure*}[htbp]
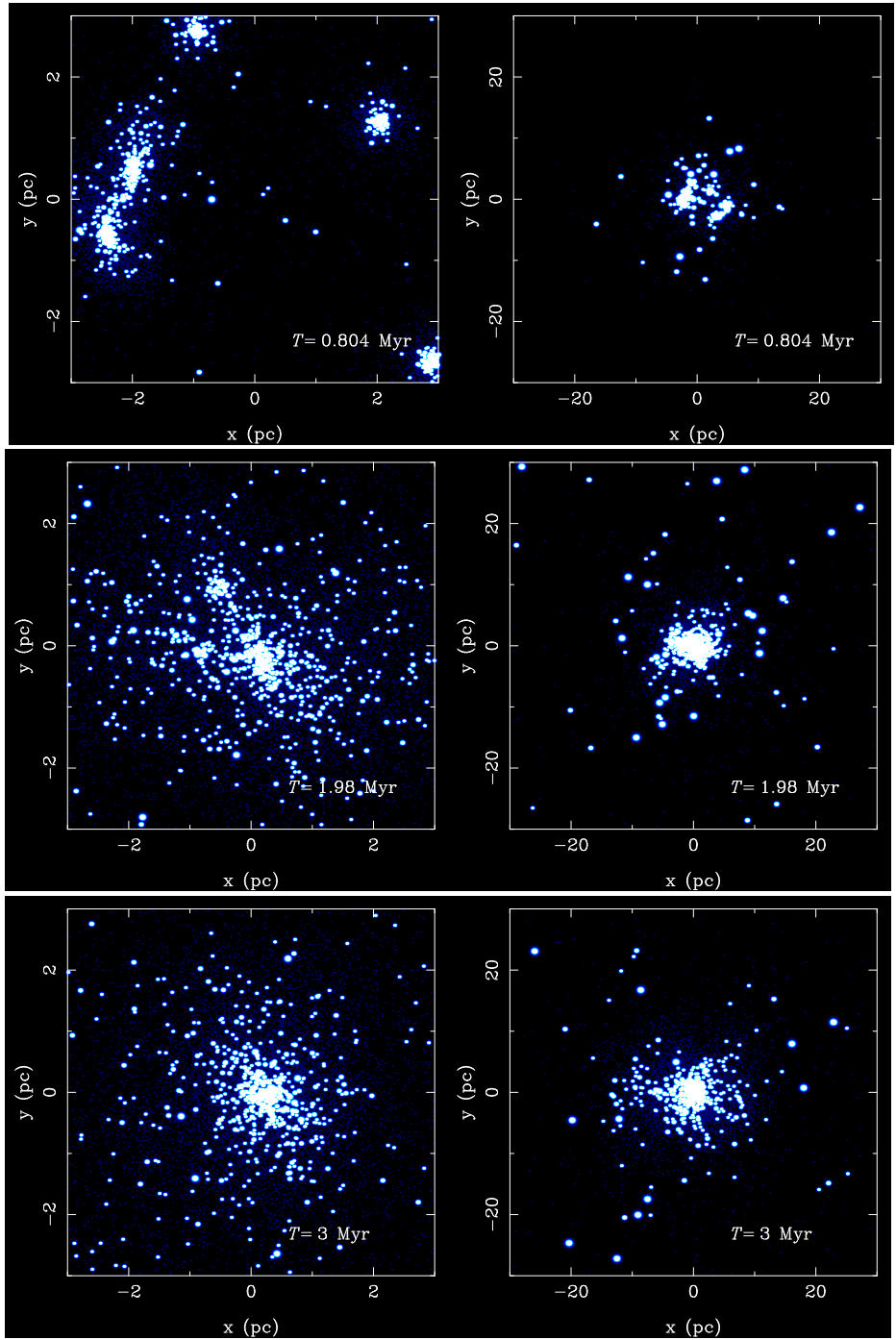

\begin{center}
\epsscale{0.8}
\plotone{f2a.eps}
\plotone{f2b.eps}
\plotone{f2c.eps}
\caption{Snapshots of model m2k8r6-1 at 0.8, 2, and 3 Myr. Left column is 
the zoom-in of the right column. \label{fig:snapshot}}
\end{center}
\end{figure*}

After 3 Myr, the sub-clusters merged to one cluster for all merger 
simulations. We present the time evolution of model m2k8r6 until 3 Myr 
in Figure \ref{fig:snapshot}. The sub-clusters hierarchically
merge and finally form one star cluster with a number of escaping 
stars. The merger remnants are almost 
spherical and look very similar to initially spherical single clusters.
However, the distribution of stars in the merged clusters are different 
from that of clusters that were formed as one.  In this section we demonstrate 
that the merger scenario is preferable for the formation of young dense 
clusters.

\subsection{Cumulative number of massive stars\label{sec:mass_dist}}

\begin{figure*}[htbp]
\begin{center}
\epsscale{1.0}
\plottwo{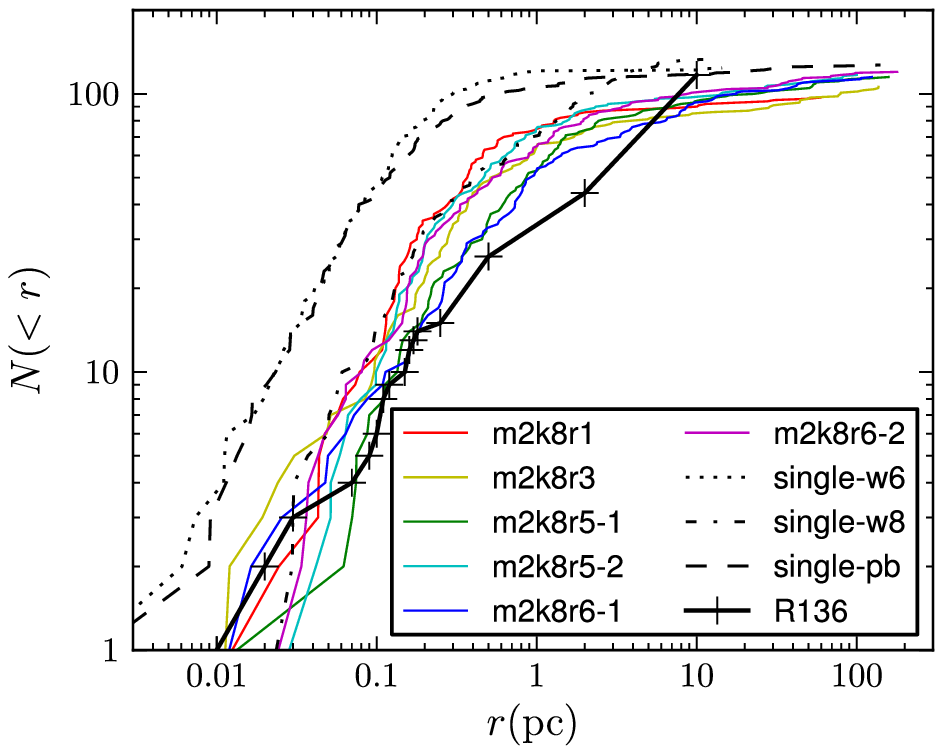}{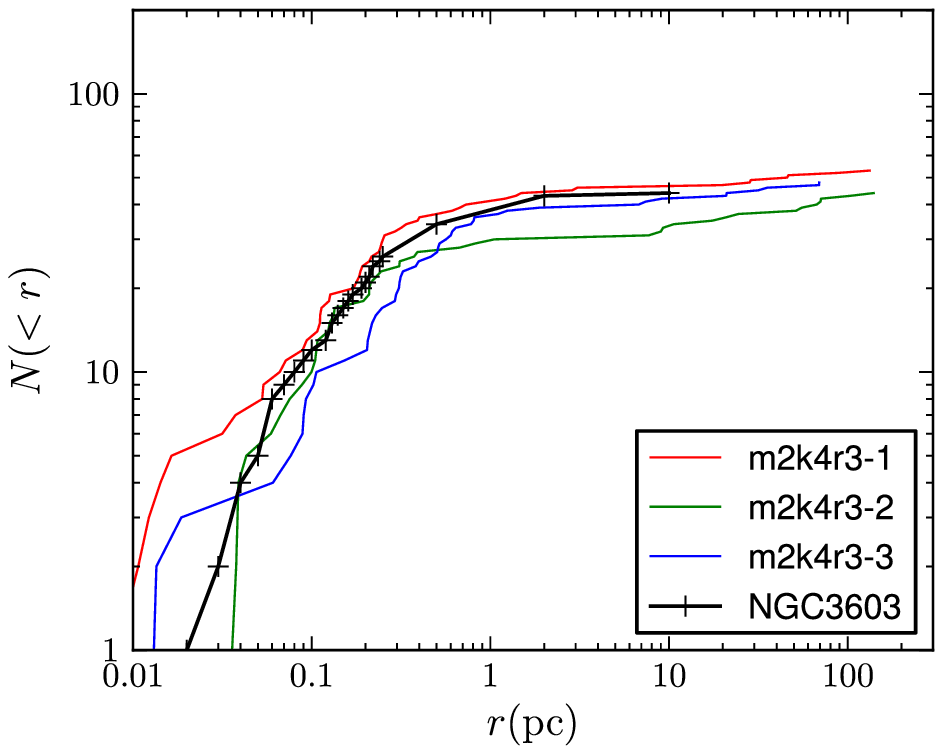}
\caption{Projected cumulative number distribution of the massive 
($>30M_{\odot}$ for simulations) stars for R136 (left) and NGC 3603
(right) models.
Colored curves show merger models and black dashed, dash-dotted, and 
dotted curves show single models. Black crosses show the observation
 \citep{1998MNRAS.296..622C}.
\label{fig:rm_R136}}
\end{center}
\end{figure*}

The projected cumulative distributions of massive 
stars as a function of the distance from the cluster center at a
cluster age of 3\,Myr are presented in Figure \ref{fig:rm_R136}.  We
adopted the density center \citep{1985ApJ...298...80C} as the cluster
center.  The black dashed and dotted curves show the results for the
single clusters, whereas the colored curves show the results for the
merger simulations. We plot massive stars with $>30 M_{\odot}$ 
for the simulations. For comparison, we overlay the observed
distribution for massive stars Wolf-Rayet (WR), spectral type O, and
some early B stars from R136 (left) and from NGC 3603 (right) 
\citep[thick black curve with crosses] {1998MNRAS.296..622C}.
Although the mass range we adopted for the simulations is 
higher than that of the observation, the fraction of plotted stars
to the total mass of the cluster is similar for both the simulation 
and the observation. 
If we assume that a Salpeter mass function between 0.3--100 $M_{\odot}$,
the fraction of O-stars ($>16M_{\odot}$) is 13.5\%, which corresponds
to the fraction of stars with $>30 M_{\odot}$ assuming a 
Salpeter mass function between 1--100 $M_{\odot}$.

In the left panel of Figure \ref{fig:rm_R136}, we find that the merger 
models and the single cluster with a relatively low central density 
(single-w8) are consistent with R136. 
We quantify the spatial distribution of the
massive stars using the radius of the inner tenth star, $r_{10}$. These 
values are $0.11\pm 0.02$, $0.087$, and $0.15$ pc for merger models,
single-w8, and R136, respectively. Model single-w8 does not 
experience core collapse sufficiently quickly to explain the high
degree of mass segregation inside the star cluster, nor does this
model produce the observed number of runaway stars (the detail 
is in section 3.2). 
On the other hand, single models with a high central density (single-w6) 
and with primordial binaries (single-pb) can produce enough runaway stars, 
but produce too a concentrated distributions of massive stars. Their values 
of $r_{10}$ are $0.034\pm 0.005$ and $0.025$ pc for single-w6 and
single-pb, respectively, and these values are much smaller than that 
of R136. We also find the agreement between merger models and  
observation for NGC 3603 (see right panel of Figure \ref{fig:rm_R136}).

We also see in the left panel of Figure \ref{fig:rm_R136} that a 
wider initial distribution of sub-clusters results in a relatively 
wider distribution of massive stars after the mergers, i.e., 
the value of $r_{10}$ is larger with the larger $r_{\rm max}$. While
$r_{10} = 0.096\pm 0.002$ pc for the models with $r_{\rm max}=$ 1--3 pc, 
$r_{10} = 0.12\pm 0.03$ pc for the models with $r_{\rm max}=$ 5--6 pc.  
The wider models can reproduce the distribution of massive stars in R136
better.  

\begin{figure}[htbp]
\begin{center}
\includegraphics[width=80mm]{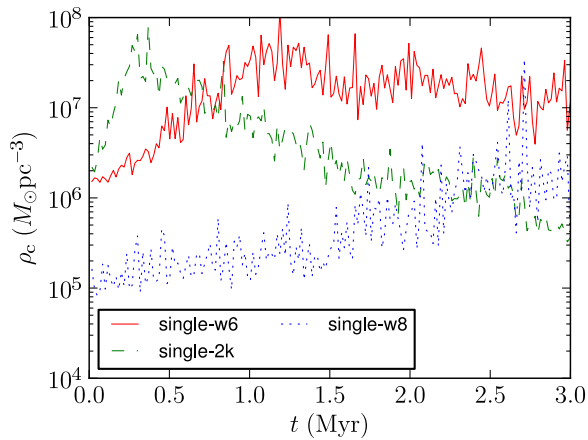}
\caption{Time evolution of core density. These are averaged on all 
runs for models single-w6 and single-2k. \label{fig:core_dens}}
\end{center}
\end{figure}

We find that the degree of dynamical evolution in each of the
sub-clusters before they merge makes this difference.  The
sub-clusters merge after they experience core collapse for the
majority of our simulations except model m2k8r1.  In the latter case,
the time scales for merger and core collapse are comparable. One would
naively expect that the more dynamically developed sub-clusters will
cause the merger product to be more centrally concentrated also, but
that is only partially true. The dynamically evolved sub-clusters
results in the lower concentration because of their shorter relaxation
time.  Figure \ref{fig:core_dens} shows the time evolution of the core
density for single clusters. The isolated sub-clusters (single-2k)
collapse at $\sim0.3$ Myr and then the core density decreases
dramatically due to the scattering of stars by hard binaries formed in
a sub-clusters before they merged. This lower core density after the
core collapse induces the lower concentration of the merger remnants.
Contrary to this, initially massive single clusters maintain their
high densities much longer.

\begin{figure*}[htbp]
\begin{center}
\epsscale{0.8}
\plotone{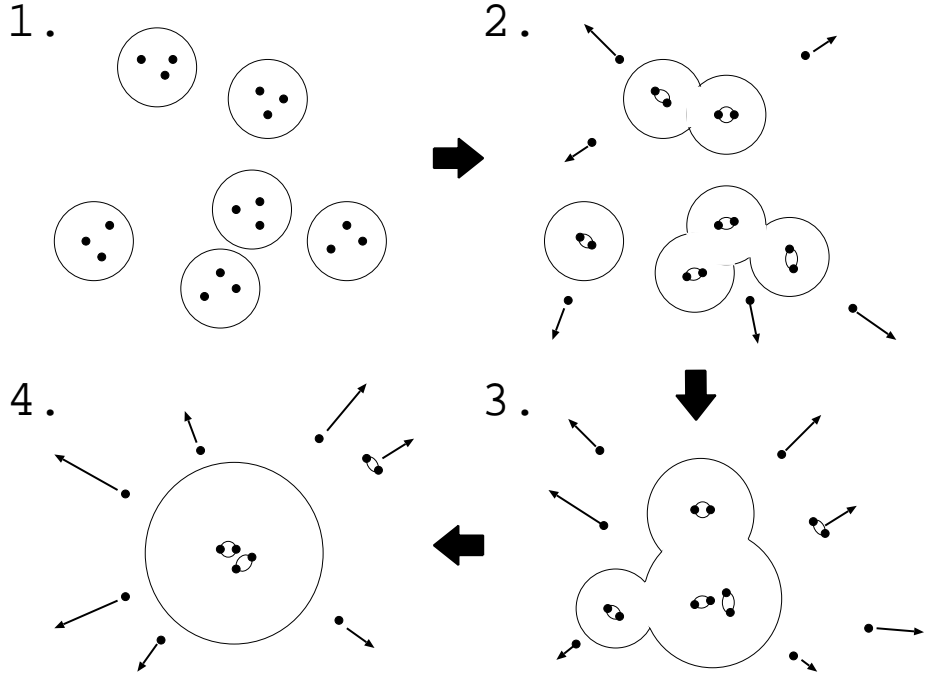}
\caption{Formation of a star cluster through mergers. Sub-clusters
  form in a region of several pc (left top). They dynamically evolve
  and experience core collapse within 1 Myr. A BB form in each cluster
  and produce runaway stars by three-body scattering (right top). The
  sub-clusters merge hierarchically and the hardest BBs
  break or kick out the other BBs
  (bottom). \label{fig:merger}}
\end{center}
\end{figure*}

The wider initial distribution of sub-clusters provides sufficient time
for them to experience core collapse and produce a BB, which ejects
massive stars from each sub-cluster (FPZ2011). The post-collapse
sub-clusters have a lower density than those in a state of core
collapse.  After the sub-clusters merge, however, most of the BBs are
broken or scattered up due to binary-binary interactions
\citep{2004MNRAS.350..615G}, and the production of massive runaways is
finally driven by one single BB. The remaining cluster starts to
evolve as an ordinary single cluster.  This explains the wider radial
profile of the more massive stars.  We schematically represent this
process in Figure \ref{fig:merger}.

\begin{figure*}[htbp]
\begin{center}
\epsscale{1.0}
\plottwo{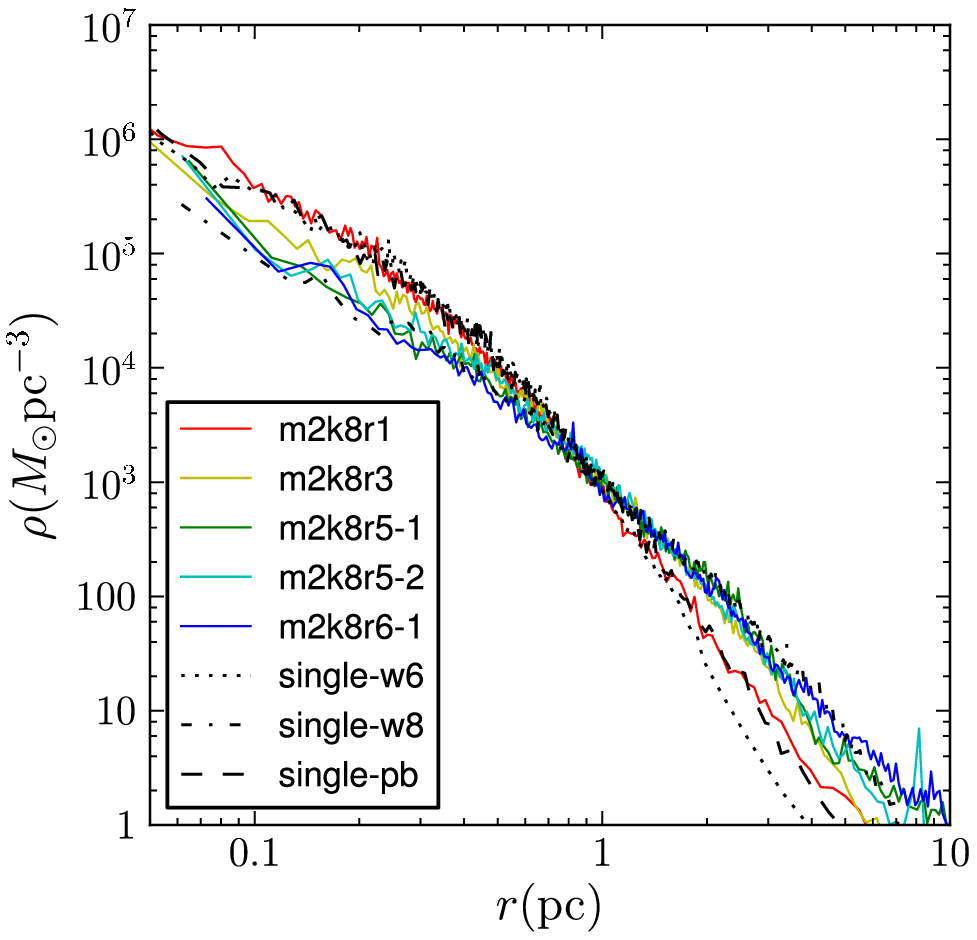}{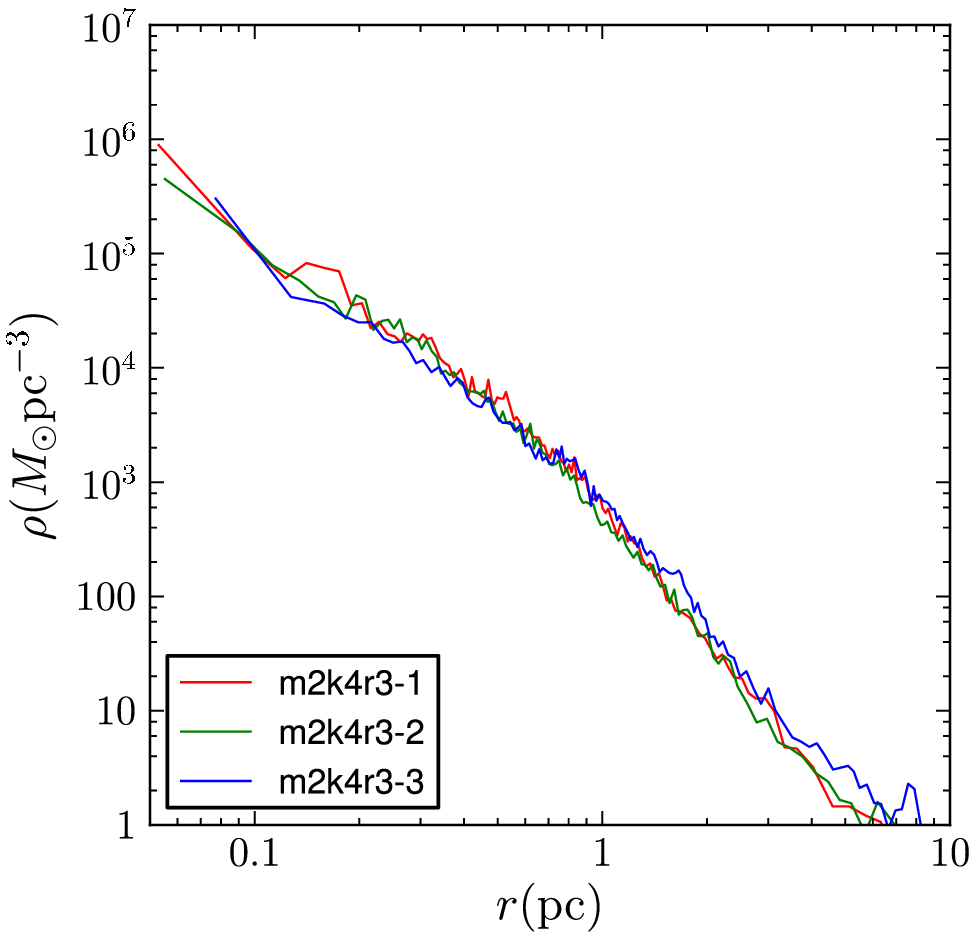}
\caption{Density profiles of R136 (8-merger and single) models at 3 Myr (left) 
and NGC 3603 (4-merger) models at 2.5 Myr (right). \label{fig:density_R136}}
\end{center}
\end{figure*}

In Figure \ref{fig:density_R136} we present the density profile of the
simulated clusters at 3 Myr. The central density of the merger
products has already been depleted except model m2k8r1.  These merger
models and the single model single-w8 show similar distributions of
massive stars and density profiles. Only the most concentrated merger
model, m2k8r1, has parameters similar to a single dense
model, single-w6.  The central density ($r\aplt 0.2$ pc) of the merger
cases with $r_{\rm max}=$5--6 pc is similar to the observed central
density of R136, 3--5$\times 10^4$ $M_{\odot}$pc$^{-3}$
\citep{2003MNRAS.338...85M}, and the central density of the
four-merger cases is comparable to the core density of NGC 3603, $\geq
6\times 10^4$ $M_{\odot}$pc$^{-3}$ \citep{2008ApJ...675.1319H}.

From the too concentrated distribution of massive stars, we can exclude 
the single high density models (single-w6 and single-pb) as viable 
models of R136. If we see only the radial distribution in the cluster, 
however, the difference between the merger and single cases is not clear.

\subsection{Mass distribution of runaway stars}
Another way to evaluate the degree of the dynamical evolution of young
star clusters is finding runaway stars around the cluster. We
demonstrated in FPZ2011 that the fraction of runaway stars around R136
and also their mass distribution are consistent with those scattered
by a binary in the core of the host cluster.

In FPZ2011, we discussed how a single cluster can explain the fraction 
of runaway stars around R136. The degree of mass segregation in our model
single-w6 is too strong compared with the observed distribution of massive 
stars (see Figure \ref{fig:rm_R136}).  On the other hand, if we assume a 
single cluster with a distribution of massive stars (single-w8) similar to 
R136, no runaway stars are produced because the relaxation time of this model 
($t_{\rm rh}=44$ Myr) is too long to experience core collapse 
within 3\,Myr (see Figure \ref{fig:core_dens}).

\begin{figure*}[htbp]
\begin{center}
\epsscale{1.0}
\plottwo{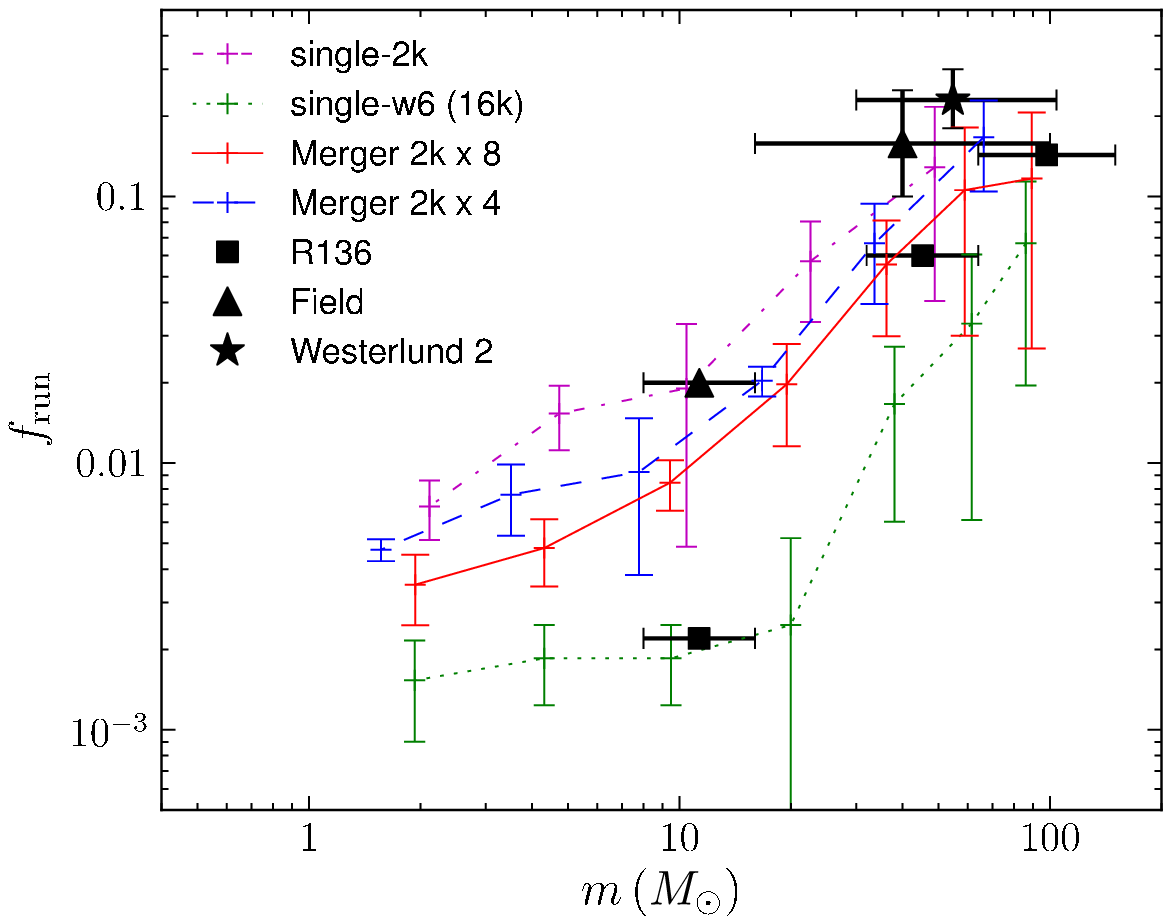}{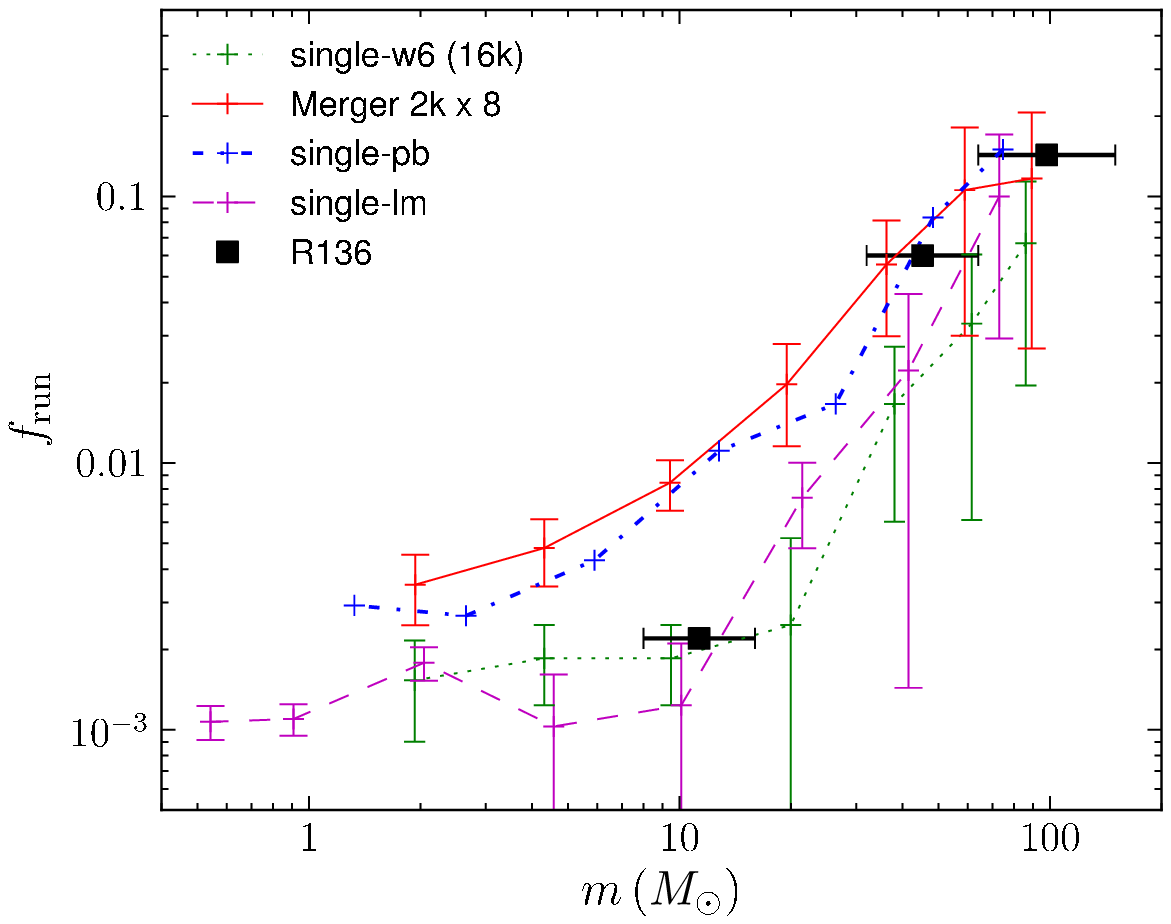}
\caption{Fraction of runaway stars ($>30$ km/s) to whole cluster as a function of mass at 3 Myr. Color curves show the results of our simulations. Left and right panels show comparisons between single and merger models and among single models, respectively.  Square and triangle symbols show the runaway fraction of R136 and the field \citep[from FPZ2011;][]{2010A&A...519A..33G,2011A&A...530L..14B,2009ApJ...707.1347A,2007ARA&A..45..481Z}. Star shows a fraction of isolated WR and early O-stars around Westerlund 2 \citep{2007ApJ...665..719T,2007A&A...463..981R,2008A&A...483..171N,2011MNRAS.416..501R}. \label{fig:esc_frac}}
\end{center}
\end{figure*}

We conclude that both the observed distributions of massive stars and
runaway stars of R136 can be reproduced if the star cluster formed
through the hierarchical merger of $\sim 8$ small sub-clusters. In
Figure \ref{fig:esc_frac} we show that $f_{\rm run}$, the fraction of
runaway stars that escape their host cluster with $>30$km/s, agrees
well with the WR and spectral type O- and early B-stars observed
around R136 (square).  The observed $f_{\rm run}$ of early B-type
stars (8--16 $M_{\odot}$) is a few times smaller than that expected
from our simulations.  Assuming that the total mass of $6.0 \times
10^4 M_{\odot}$ \citep{2010ARA&A..48..431P} with a Salpeter mass
function between 1 and 100 $M_{\odot}$ and that the runaway fraction
obtained from our 8-merger models, we predict that there are $7 \pm 2$
early-B runaways still in the vicinity of R136. We also expect that
$\sim 6$ late-O (16--32 $M_{\odot}$) and $\sim 11$ early-O/WR ($>32
M_{\odot}$) runaway stars exist around R136.

\begin{figure}[htbp]
\begin{center}
\epsscale{0.8}
\plotone{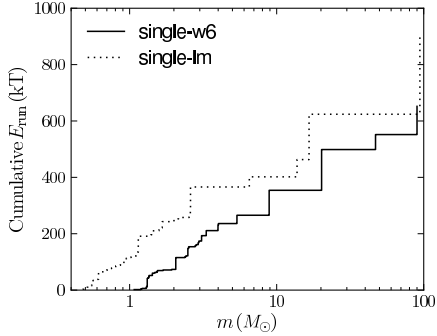}
\caption{Kinetic energy of runaway stars as a cumulative function of mass. The energy is normalized by kT of each cluster. Note that the value of kT for model single-w6 is twice as large as that of single-lm because of the difference of their mean masses.\label{fig:E_run}}
\end{center}
\end{figure}

The lower-mass limit of the IMF does not affect on the fraction
of runaway stars (see the right panel of Figure \ref{fig:esc_frac}) because
the three-body scattering by a binary in the cluster core is driven 
by massive stars gathering in the cluster core due to the mass segregation. 
Even if we change the lower-mass limit of the IMF, low-mass 
stars do not interact much frequently with the binary in the core. 
In Figure 
\ref{fig:E_run} we show the kinetic energy of runaway stars as a
cumulative function of mass, which corresponds to the energy 
extracted from the binary by three-body scatterings. 
The contribution from low-mass stars is much smaller than that from
massive stars. The core-collapse time is not sensitive to the
lower-mass limit either, although the number of particles increases 
with a smaller lower-mass limit.
From the results of $N$-body simulations, the core-collapse time, 
$t_{\rm cc}$, follows an equation such as
$t_{\rm cc} \propto (m_{\rm max}/\langle m\rangle)^{-1.3} t_{\rm rh}$ 
\citep{2004ApJ...604..632G}, where $m_{\rm max}$ and $\langle m\rangle$ are 
lower-mass limit and mean mass, respectively.
Since $\langle m\rangle \propto 1/N$ and the relaxation time is roughly 
proportional to $N$, the core-collapse time depends on $N^{-0.3}$.
The lower-mass limit of the IMF, however, changes the mass where 
the fraction of runaway stars starts increase because only stars 
which are more massive than this limit mass can sink to the cluster 
center \citep{2008ApJ...686.1082F}.  A smaller value of the lower-mass 
limit decrease this mass \citep[see Figure \ref{fig:esc_frac} and]
[in which they assumed an IMF between 
0.3 and 150 $M_{\odot}$ and obtained very similar results to ours.]
{2012ApJ...746...15B}.  The limit mass is 13 and 
9 $M_{\odot}$ for our models with a lower-mass limit of 1 and 
0.47 $M_{\odot}$, respectively. 

The consistency between the observed number of runaways and the
density profile for R136 gives us confidence in our prediction of a
relative fraction of runaway stars around NGC 3603 and other clusters.
As we 
discussed before, the fraction of runaway stars does not 
depends on the lower-mass limit of the IMF. 
Therefore, we adopted both 0.3 and 1 $M_{\odot}$ as the 
lower-mass limit and assumed the fraction of runaway stars obtained
from our simulations a the lower-mass limit of 1 $M_{\odot}$. 
Since the number of massive stars assuming a lower-mass limit of 0.3 
$M_{\odot}$ is $\sim$60\% of that in the case of 1 $M_{\odot}$, 
the expected number of runaway stars also decrease $\sim$60\%. 
Adopting a Salpeter mass function between 1 $M_{\odot}$ and 100
$M_{\odot}$ and a total cluster mass of $\sim 1.7 \times 10^4
M_{\odot}$ \citep{2010ApJ...716L..90R}, we predict that $\sim 7$ O/WR
stars ($>16 M_{\odot}$) have been ejected from NGC 3603. If we assume
a lower-mass limit of 0.3 $M_{\odot}$, we predict $\sim 4$ O/WR runaway 
stars. We summarize
the expected number of OB runaway stars around young star clusters in
table \ref{tb:runaway}. We assumed 8-merger models for R136 and 
Westerlund 1 and 4-merger models for NGC 3603 and Westerlund 2.

\begin{table*}[htbp]
\begin{center}
\caption{Expected runaway stars.\label{tb:runaway}}
\begin{tabular}{cccccccc}\hline
Name    & $M_{\rm cl} (M_{\odot})$ & Ref.& Model & $m_{\rm min} (M_{\odot})$ & early-B  & late-O  & early-O/WR \\ \hline\hline
R136         & $6.0 \times 10^4$ & 1 & 8-merger & 1 & $6.7\pm 2.4$ & $6.4\pm 3.0$ &  $11\pm 6$\\ 
 & & &  & 0.3 & $4.9\pm 1.5$ & $3.9\pm 1.8$ &  $6.5\pm 3.5$\\ 
NGC 3603      & $1.7 \times 10^4$ & 2 & 4-merger & 1 & $2.9\pm 2.2$ & $2.1\pm 0.6$ &  $4.8\pm 1.5$\\
 & & &  & 0.3 & $1.8\pm 1.3$ & $1.3\pm 0.3$ &  $2.9\pm 0.9$\\
Westerlund 1 & $6.3 \times 10^4$ & 1 & 8-merger & 1 & $7.0\pm 2.5$ & $6.8\pm 3.2$ &  $11\pm 6$\\ 
 & & &  & 0.3 & $4.2\pm 1.5$ & $4.1\pm 1.8$ &  $6.8\pm 3.7$\\ 
Westerlund 2 & $10^4$            & 1 & 4-merger & 1 & $1.7\pm 1.3$ & $1.2\pm 0.3$ &  $2.8\pm 0.9$\\
 & & &  & 0.3 & $1.0\pm 0.8$ & $0.8\pm 0.2$ &  $1.7\pm 0.8$\\
\hline
\end{tabular}
\tablecomments{References: 1. \citet{2010ARA&A..48..431P}, 2. \citet{2010ApJ...716L..90R}. The lower-mass limit is indicated as $m_{\rm min}$. }
\end{center}
\end{table*}

\subsection{Single cluster with primordial binaries}
Since the binary fraction of massive stars in star clusters is very
high (20--80\%) \citep[][and references therein]{2007ARA&A..45..481Z},
and binary-binary interactions are common and can effectively produce
runaways
\citep{1988AJ.....96..222L,1992MNRAS.255..423C,2004MNRAS.350..615G}, a
fraction of primordial binaries and their distribution in binding
energies are expected to affect the fraction of runaway stars.  We
performed a simulation for a single cluster model with primordial
binaries (single-pb) in order to quantify the effect of binaries.  The
radial distribution of massive stars (Figure \ref{fig:rm_R136}) and
the density profile (Figure \ref{fig:density_R136}) are very similar
to the single case with the same initial density distribution
(single-w6).  The primordial binaries do not solve the problems that
clusters require too high a core density and too concentrated a
distribution of massive stars to form a sufficient amount of runaway
stars; primordial binaries therefore do not help in reproducing the
spatial distribution of the stars in R136.
 
The density profile of the cluster is not affected by the presence of
primordial binaries, but the fraction of the runaway stars is higher
compared to those produced in single clusters, and similar to those
produced in merged cluster (see Figure \ref{fig:esc_frac}). This
fraction is consistent with the simulations performed by
\citet{2012ApJ...746...15B}.

\subsection{Comparison with Westerlund 1}

\begin{figure}[htbp]
\begin{center}
\epsscale{0.8}
\plotone{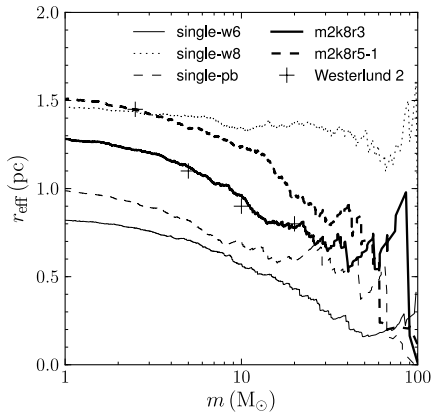}
\caption{Effective cumulative radius for stars in 3 pc (in projection) at 3.5 Myr. Data points of Westerlund 2 are from \citet{2011MNRAS.412.2469G}. \label{fig:eff_rad}}
\end{center}
\end{figure}

Westerlund 1 is another young massive star cluster in the Milky Way
with a mass comparable to R136, but with an age of 3--6 Myr
\citep{2005A&A...434..949C,2011MNRAS.412.2469G}.  The relaxation time
estimated from its current mass and size is $\sim 130$ Myr, but the
cluster is mass-segregated \citep{2011MNRAS.412.2469G}.  Furthermore,
Westerlund 1 is highly asymmetric, which could indicate that it had
experienced mergers. Unfortunately, there is no data of runaway stars
around this cluster, and little data is available about
its internal kinematics. However, the effective cumulative radius:
\begin{eqnarray}
r_{\rm eff}(m) = \sqrt{\frac{\sum_{m_{i}>m}^i r_{i}^2}{i}},
\end{eqnarray}
has been observed \citep{2011MNRAS.412.2469G}. Here, $r_i$ is the
distance from the cluster center in projection for $i$-th massive
stars with $m_{i}$. This is the geometric-averaged distance from the
cluster center for stars with mass $m_i > m$. We calculated $r_{\rm
  eff}(m)$ for our simulations in order to compare it with the
observation.  We took into account stars in a radius of 3\,pc to
remove the effect of runaway stars. This radius is the same as the
observed region, $\sim$ 2.3\,pc. The results are shown in Figure
\ref{fig:eff_rad}.  The smaller effective radius for more massive
stars indicates mass segregation. With simulation m2k8r3, which has a
density profile which is consistent to the observed cluster R136, we
quantify the observation to $r_{\rm eff}\sim 0.8$ pc at the minimum
radius at massive end ($\sim 20 M_{\odot}$) and $r_{\rm eff}\sim 1.5$
pc for all stars down to 2.5 $M_{\odot}$ \citep[see also Figure 11 of
][]{2011MNRAS.412.2469G}.  The effective radius for stars with 
$\apgt 30 M_{\odot}$ is larger than than that at around 
20--30$M_{\odot}$ in both the simulation and the observation.
In the simulation the larger radius comes from some off-center massive 
stars, which are scattered from the cluster center. 
In contrast to the merger models, the single model single-w8,
which initially has properties similar to of that of the current
Westerlund 1, has not have sufficient time to evolve dynamically. From
these results, we consider that Westerlund 1 is also a young massive
cluster which is forming via hierarchical merger. Since Westerlund 1
(3--6 Myr) is slightly older than R136 ($\sim 3$Myr) but is still 
undergoing mergers, we expect that the sub-clusters of Westerlund 1 
were born in a larger region than those of R136. We expect
that based on the simulation, the numbers of runaway stars of
Westerlund 1 are $7.0\pm 2.5$ ($4.2\pm 1.5$), $6.8\pm 3.2$ ($4.1\pm 1.8$), 
and $11\pm 6$ ($6.8\pm 3.7$) for
early-B, late-O, and early-O/WR stars assuming a Salpeter IMF with 1--100
$M_{\odot}$ (0.3--100$M_{\odot}$), respectively (see also table
\ref{tb:runaway}).

\section{Distribution of Collided Stars}
Recent observations have identified several very massive stars with an
estimated zero-age main-sequence mass of $\apgt 150 M_{\odot}$ in R136
and NGC 3603 \citep{2010MNRAS.408..731C}. Because collisions between
stars easily happen in a dense cluster \citep{1999A&A...348..117P},
such very massive stars are a natural by-product of the dynamical
evolution of a dense star cluster.

\begin{figure*}[htbp]
\begin{center}
\epsscale{1.0}
\plottwo{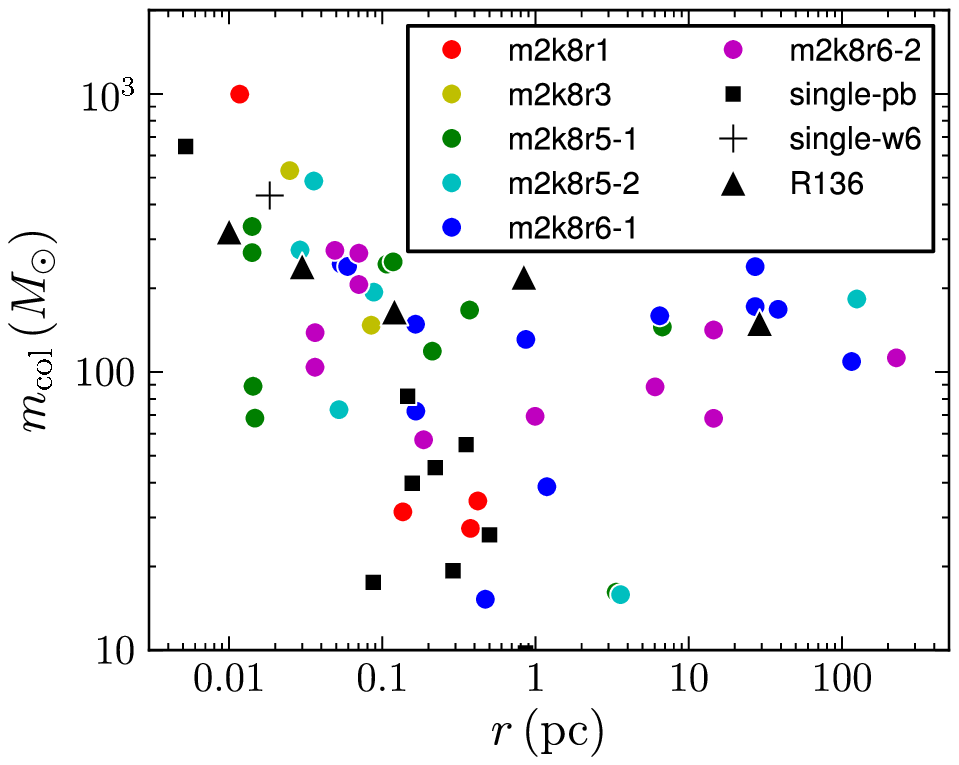}{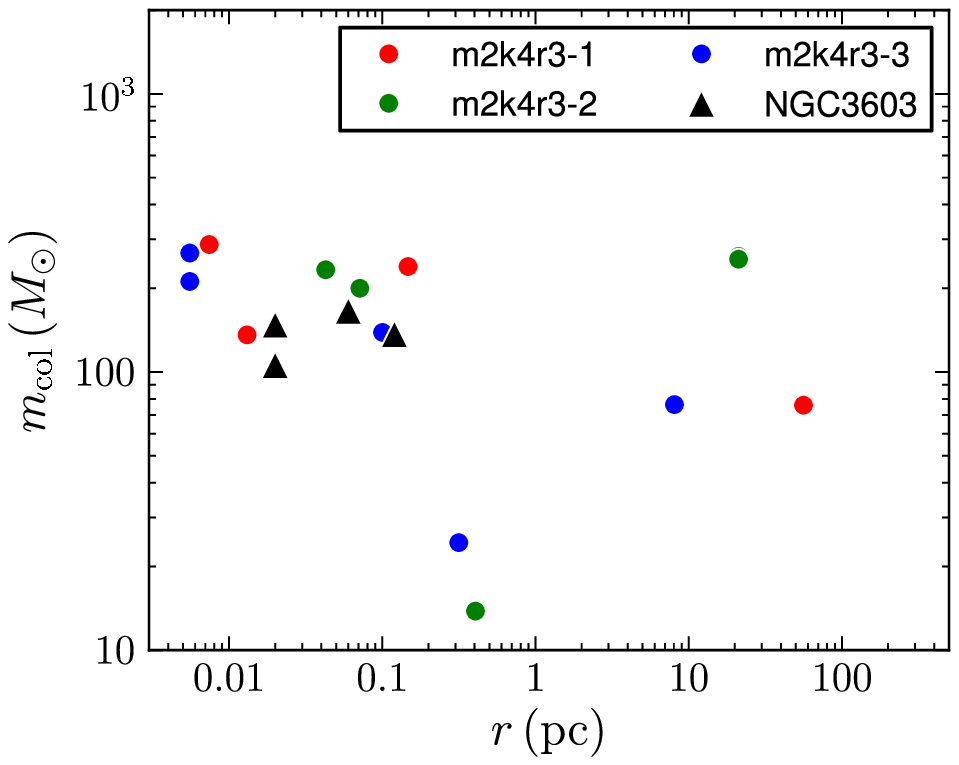}
\caption{Radial and mass distribution of collided stars at 3 Myr for
  8-merger models (left) and 2.5 Myr for 4-merger models (right).
  Circles show merged cluster cases, whereas squares show a single
  case with primordial binaries. The cross shows the averaged value of
  single cluster models without primordial binaries (single-w6). In
  the case of single clusters, only one runaway collision star grows
  in a cluster. Black triangles show very massive stars
  ($>100M_{\odot}$) found in R136 and NGC 3603
  \citep{2010MNRAS.408..731C,2011A&A...530L..14B}. The right top green
  point in the right panel is a binary which have very similar
  masses.\label{fig:merged_star}}
\end{center}
\end{figure*}

In Fig.\,\ref{fig:merged_star} we present the radial distribution and
the mass distribution of stars that experienced collisions in our
simulations. We adopted for adopting the maximum masses a star reaches
during its lifetime rather than the mass at the end of the simulation,
because the latter strongly depends on the mass-loss rate. For
comparison, we also indicate the very massive stars observed in R136
and NGC 3603 \citep{2010MNRAS.408..731C,2011A&A...530L..14B}.  In the
left panel of Figure \ref{fig:merged_star}, we can see the effect of
the mass segregation within $\sim 1$ pc, while we find stars escaping
at distances greater than 10 pc. The escaping massive stars formed in
the merger models with $r_{\rm max}=$ 5--6 pc. The maximum mass that
was produced in these simulations was around $240M_{\odot}$, and three
of these stars are ejected from the cluster as single runaways. Their
velocities are similar, $\sim 60$ km/s in three dimensions, which
corresponds to $\sim 35$ km/s in one dimension. This mass and velocity
are very similar to those of VFTS 682, which is located at $\sim 30$
pc from R136 in projection and seems to be escaping from R136 with
one-dimensional velocity of $\sim 30$ km/s
\citep{2011A&A...530L..14B}. With a current mass of VFTS 682 is $\sim
150 M_{\odot}$ its initial mass would $\aplt 210 M_{\odot}$
\citep{2011A&A...530L..14B}.

\begin{figure*}[htbp]
\begin{center}
\epsscale{1.0}
\plottwo{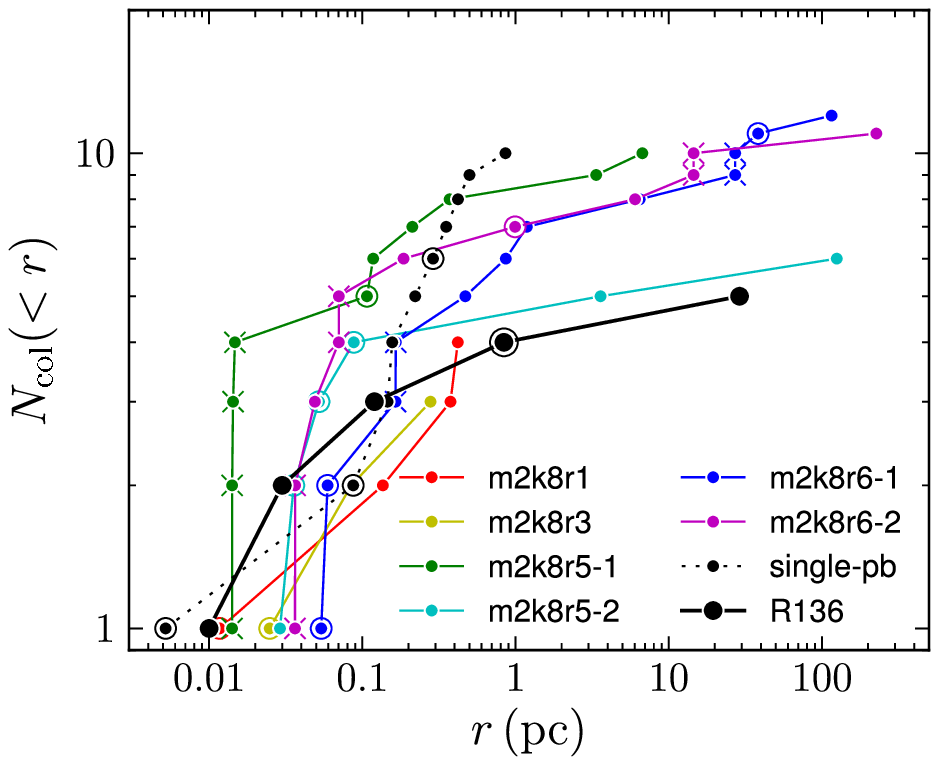}{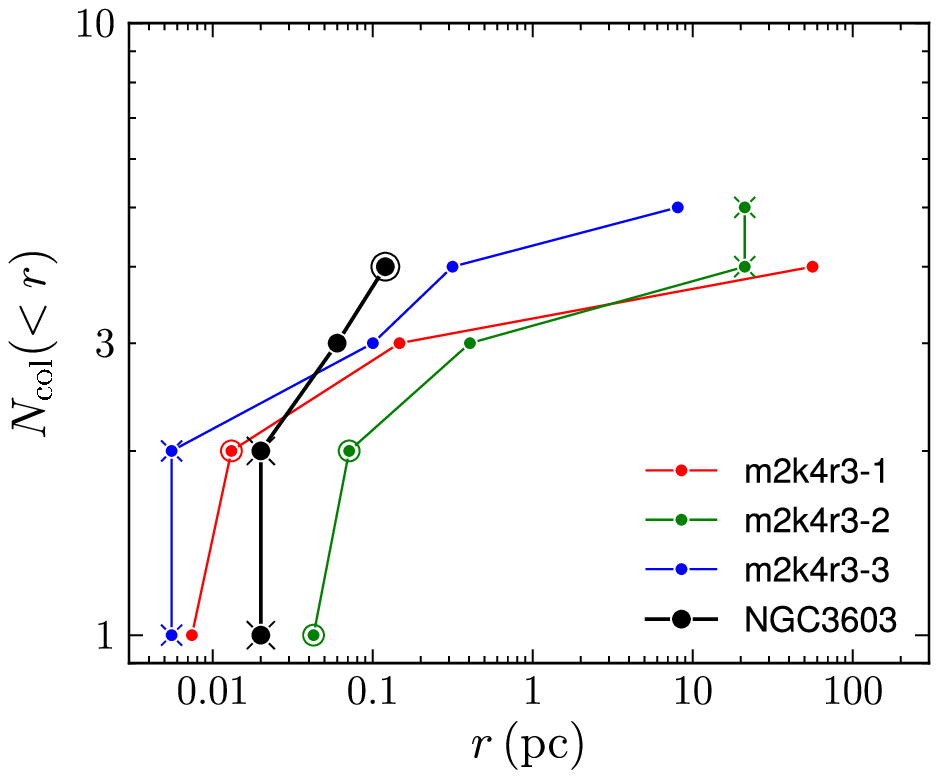}
\caption{Cumulative radial distribution of stars which experienced
  collisions and with $>100 M_{\odot}$ at 3 Myr for R136 models (left)
  and 2.5 Myr for NGC 3603 models (right). Crosses indicate binaries
  whose companions are also collided stars (two crosses at the same
  radius are a pair). Surrounded circles indicate binaries whose
  companions are not collided stars.  \label{fig:nr_merged_star}}
\end{center}
\end{figure*}

Such very massive stars have been found also inside of R136.  There
are four massive stars with initial masses of 165--320 $M_{\odot}$ in
R136 \citep{2010MNRAS.408..731C}.  Three of four massive stars in R136
are located in the inner most 0.1 pc and the other one is $\sim 1$pc
(in projection) from the cluster center.  In our simulations, we found
3--4 very massive stars in the cluster center in the case of merged
clusters with $r_{\rm max}=$ 5--6 pc.  Model m2k8r6-1 evolves towards
a distribution very similar to R136. In Figure
\ref{fig:nr_merged_star}, we present the radial distribution of merger
products with $>100 M_{\odot}$.

We compared the distribution of massive stars in NGC 3603 with the
simulations. There are four very massive stars with initial masses of
105--170 $M_{\odot}$ in NGC 3603, and two of them (A1a and A1b) form a
binary \citep{2010MNRAS.408..731C}. We found 2--3 collided stars which
exceed 100 $M_{\odot}$ in each run. Some of the stellar merger
products form binaries. We find a massive binary composed of stars
that experienced a collision in model m2k4r3-3 and one composed of a
collided star and a $\sim 100 M_{\odot}$ normal star in model
m2k4r3-1. We speculate, based on our simulation results, that massive
($>100M_{\odot}$) star could have been ejected from NGC 3603 (see
right panels of figures \ref{fig:merged_star} and
\ref{fig:nr_merged_star}), and could still roam the LMC.

The merger simulations that started with a relatively large $r_{\rm
  max}$ have mass and density profiles that are consistent with the
observations, but the single clusters and merger simulations with
small $r_{\rm max}$ are unable to reproduce the observations.  In the
latter cases, one very massive star $\apgt 400M_{\odot}$ forms in the
cluster as a result of runaway collisions (see Figure
\ref{fig:merged_star}). However, there is no observational evidence of
such a massive star in R136 or NGC 3603. 

In the simulations with primordial binaries (single-pb), no massive
runaway star which experienced collisions formed, whereas 
\citet{2012ApJ...746...15B} found some
collided runaway stars in their $N$-body simulation models of R136 for
a single cluster with primordial binaries. In our simulation, most of 
the massive stars collided into the most massive star via runaway 
collisions as is the case without primordial binaries,
which is consistent with \cite{1999A&A...348..117P}. We argue that the
difference between our results and those presented in
\citet{2012ApJ...746...15B} originate from the initial conditions.
The major differences are the orbital period distribution of
primordial binaries and their assumption that all binaries are born in
the cluster core.  We adopted the same binding energy for all binaries
(orbital periods for massive binaries are $\sim 3\times10^3$--$10^6$
days) and distribute them in the same way in the cluster as the single
stars.  On the other hand, \citet{2012ApJ...746...15B} chose shorter
periods of $0.5 < \log_{10} (P/{\rm days}) < 4$. They employed initial
mass segregation; they initialized the heaviest stars in the cluster
core. The combination of relatively tight binaries and a high degree
of mass segregation stimulate strong dynamical encounters which may in
their case have resulted in the stimulated formation of a massive
runaway in the simulations with primordial binaries.

In Figure \ref{fig:nr_merged_star} we show that about half the stars
that experienced one or more collisions are accompanied by another
star. In simulations with merging clusters, one BB forms in each
sub-cluster before it merges. The BBs in the sub-cluster centers
participate in colliding encounters due to which they become more
massive. After the sub-clusters merge, one of the BBs can be
dynamically scattered and escapes from the cluster. We discuss
binaries in section \ref{sec:binary}.

Stars that experience a collision are candidates for rapid rotation 
\citep{2005ApJ...623..302F}.
Several rapidly rotating massive stars have been found in recent
observations, and possibly this rotation
is the result of the earlier collision history of the star.  The star
VFTS 682, has a rotation velocity $> 200$ km/s
\citep{2011A&A...530L..14B}, and the $>20 M_{\odot}$ VFTS 102, which
is a runaway with $\sim 40$ km/s located in the 30 Doradus region, is
also a rapid rotator ($> 500$ km/s) \citep{2011ApJ...743L..22D}.  The
rapid rotation of these stars could be explained by an earlier episode
of mass transfer in a binary system or by collision with another star.
Either of these probably occurred in the cluster center, from which 
it can easily have been ejected.  In our
simulations we find several rotating stars inside the model that
reproduces the observations of R136 and a few are ejected as runaways
(see figures \ref{fig:merged_star} and \ref{fig:nr_merged_star}).

\section{Distribution of Binaries\label{sec:binary}}

\begin{figure}[htbp]
\begin{center}
\epsscale{1.0}
\plotone{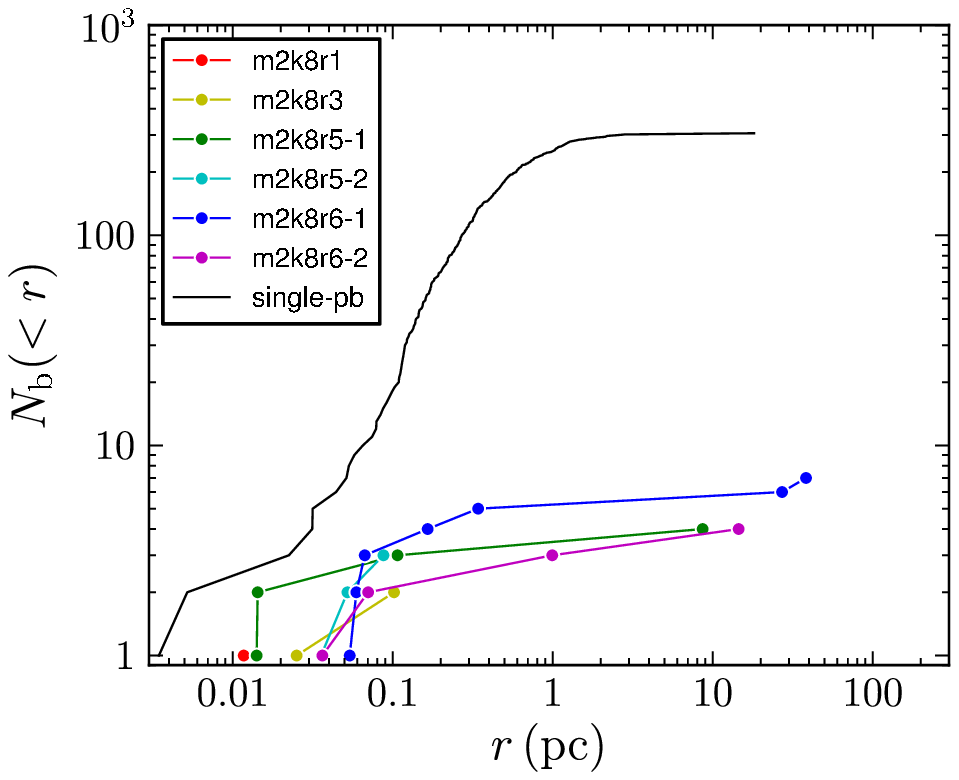}
\caption{Cumulative radial distribution of hard binaries at 3 Myr. Colored 
points show merger models, and the black curve shows a single model with primordial 
binaries. \label{fig:nr_binary}}
\end{center}
\end{figure}

In Figure \ref{fig:nr_binary} we show the radial distribution of hard
binaries (the binding energy is more than 1kT) at 3 Myr.  In the
simulations with primordial binaries, we find 306 such binaries at an
age of 3 Myr, which is $\sim 60\%$ of the primordial binary
population. The majority of these binaries are primordial.  

\begin{figure}[htbp]
\begin{center}
\epsscale{1.0}
\plotone{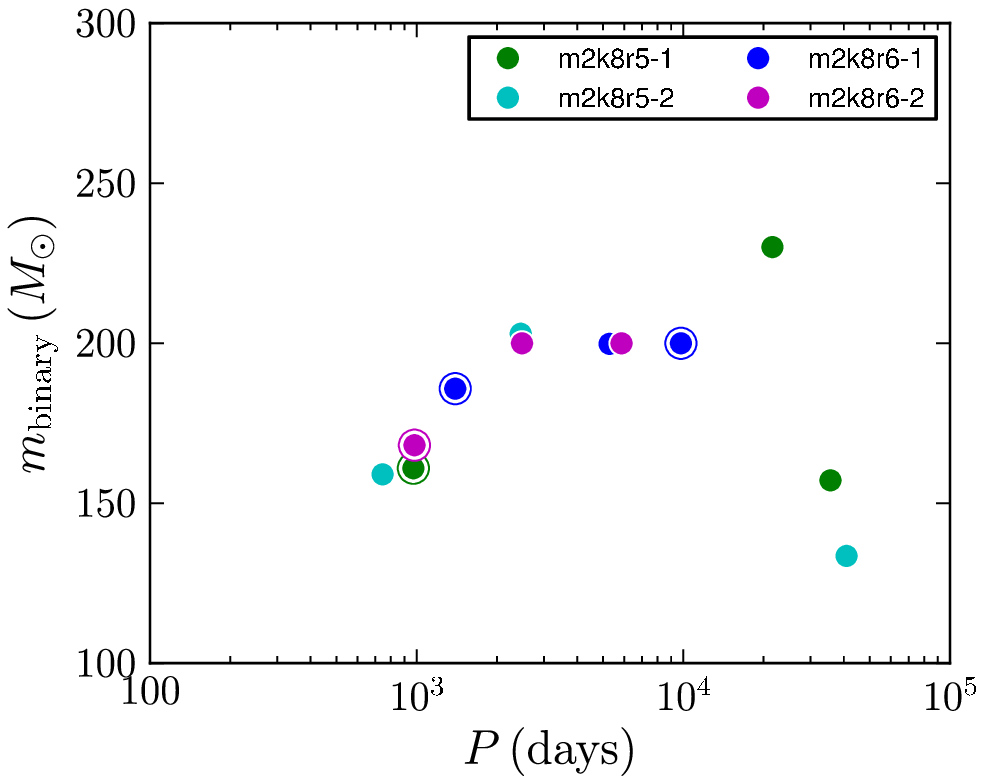}
\caption{Orbital periods and total masses of binaries at 3 Myr. Points with circles indicate binaries which are located at $>5$ pc from the cluster center.
\label{fig:binary}}
\end{center}
\end{figure}

In the case of a single cluster with/without primordial binaries and 
merged clusters with $r_{\rm max}=$ 1--3 pc, there is generally one 
BB in the cluster center.
In the case of a merged cluster with $r_{\rm max}=$ 5--6 pc, however,
we found 3--7 BBs. Furthermore, we found two escaping binaries
in model m2k8r6-1. They are located at 38 and 27 pc from the cluster
center and their velocities are 24 km/s and 16 km/s, which are
slightly slower than the velocity to be defined as runaway stars, but
these are unbound. We also found off-center binaries in two of our
merger models (models m2k8r5-1 and m2k8r6-2). They are located at 36
and 8 pc from the cluster center, but their velocity is $\sim 10$ km/s
and they are bound to the cluster. These binaries are remnants of the
BBs that formed in sub-clusters and were kicked out after their host
clusters merged.  Therefore, they are very massive with a maximum mass
$\sim 240 M_{\odot}$.  Figure \ref{fig:binary}
shows the orbital periods and the total current masses of the binaries
obtained from the simulations.  The orbital period of the massive
binaries located farther than 10 pc (marked by circle in the figure)
is $\sim 500$ days. R145, which is a massive binary (300 + 125
$M_{\odot}$) located at $\sim 20$ pc in projection from R136 and 
of which orbital period is 158.8 days
\citep{2009MNRAS.395..823S,1984ApJ...279..578F}, has similar
characteristics to these escaping binaries.

\section{Summary}
We performed simulations of merging and single star clusters
and compared the radial distribution of massive stars and the 
mass distribution of runaway stars to observations. 
We found that clusters that assembled from the hierarchical merging 
of multiple sub-clusters can explain the characteristics 
(mass distribution, mass segregation, runaway stars, massive stars, 
and massive binaries) of young dense clusters such as R136 in the 
LMC and NGC 3603, Westerlund 1, and 2 in the Milky Way.

We have set-up our simulations in such a way that the sub-clusters
have a sufficiently short relaxation time that they experience core
collapse before they merge and as a consequence the large merged
cluster is in a state of core collapse when it forms. This results in
the mass segregation of the merger remnants.  Each sub-cluster
experiences core collapse which leads to a collision runaway and the
consequent production of a very massive ($>100M_{\odot}$) star in the 
process. After the clusters have merged the BBs gather in the center of
the merged cluster. The majority of them are ionized or ejected from
the cluster out by binary-binary, in fact BB-BB, interactions.  These
processes lead to the formation of very massive 
stars, consistent with those observed in R136 and NGC 3603, massive
runaway stars such as VFTS 682, and massive binaries in the periphery
of young cluster like R145.  The post-core-collapse evolution of the
sub-clusters drives the expansion of the core by the ejection of a
large number of runaway stars.  The distribution of massive stars and
the core density of the star clusters which formed via hierarchical
mergers are consistent with those of R136 and NGC 3603.

The initial conditions of our simulations are constructed such that
core collapse occurs at an age of $\sim$0.3\,Myr. This fine tuning is
required in order to explain the coexistence of a rich population of
runaways, the presence of a BB in the cluster center (or outside) and
the relatively low density of the observed clusters.  Observing a star
cluster in a state of cluster collapse at an age $\aplt 1$\,Myr would
provide a confirmation for formation model for massive clusters and
the co-production of runaway stars and extremely massive single stars.

If we adopt single clusters with initial conditions for which the
distribution of the massive stars is consistent with observations, our
simulated clusters are insufficiently dense to experience core
collapse within a few Myr and to form runaway stars within the age of
the observed clusters. On the other hand, initial single clusters that
are sufficiently dense to experience core collapse within 3\,Myr can
produce a sufficiently large number of runaways compared to the
observed number found around R136, but they are considerably more
concentrated than the observed clusters. Single clusters with
primordial binaries mediates the number of runaway stars, but also
lead to a too concentrated density profile at later time.

Westerlund 1 and 2 are also candidates of merged clusters, in
particular because they appear to be rather clumpy and they are
strongly mass segregated \citep{2011MNRAS.412.2469G,
  2007A&A...466..137A}.  The degree of mass segregation in our merger
simulations is consistent with the observed cumulative effective
radius of Westerlund 1.  The fraction of runaway stars in the vicinity
of Westerlund 2 is consistent with the result of our simulations in
which multiple clusters merge to one.

Our results support the claim that some young dense clusters, such as
R136, NGC 3603, Westerlund 1, and 2, formed via mergers of smaller
clusters.  We further argue that the small sub-clusters which are best
suited to reproduce the observed characteristics of the young dense
clusters are also the building blocks for star clusters elsewhere in
the Milky Way, and are therefore responsible for the production of O
and B runaway stars in the Galactic disk.  The typical scale of the
cluster is $M\sim 6000 M_{\odot}$, and the fraction of runaway stars
among O-type stars is about 10--20\%, and drops to a few percent for
B-type stars (FPZ2011).  Clusters more massive than those studied here
may also have formed through the merging of multiple smaller
sub-clusters. We argue that the fundamental building-block cluster
with which we are able to explain the density profile of observed star
clusters and at the same time the rich population of massive early B
and O stars in the vicinity has a mass of $\aplt 10\,000 M_\odot$.

\acknowledgments We thank Keigo Nitadori for developing the code,
Masaki Iwasawa for fruitful discussions, and Alexander Rimoldi for
careful reading.  We also thank the anonymous referee for useful
comments.  This work was supported by Research Fellowship of the Japan
Society for the Promotion of Science (JSPS) for Young Scientists, the
Netherlands Research Council NWO (grants VICI [\#639.073.803], AMUSE
[\#614.061.608] and LGM [\# 612.071.503]), the Netherlands Research
School for Astronomy (NOVA), and HPCI Strategic Program Field 5 `The
origin of matter and the universe'.  Numerical computations were
carried out on the Cray XT4 at the Center for Computational
Astrophysics (CfCA) of the National Astronomical Observatory of Japan
and the Little Green Machine at Leiden University.

\bibliographystyle{apj}
\bibliography{apj-jour,reference}

\end{document}